\documentclass[aps,prd,preprint,groupedaddress]{revtex4-1}

\usepackage{graphicx}
\usepackage{amsmath}
\usepackage{amsfonts}      
\usepackage{hyperref}     
\usepackage{multirow}
\usepackage{xcolor}
\usepackage{slashed}
\usepackage{makecell}
\allowdisplaybreaks

\newcommand{\nn}{\nonumber\\}

\begin{document}
\pagenumbering{gobble}
\title{$D^*D^*\pi$ and $B^*B^*\pi$ couplings from light-cone sum rules}

\author{Chao Wang}
\email[]{chaowang@nankai.edu.cn}
\affiliation{ \small Faculty of Mathematics and Physics, Huai'an University, Huaian 223001,  China}

\date{\today}

\begin{abstract}
We revisit the calculation of the strong couplings $D^{*}D^{*}\pi$ and $B^{*}B^{*}\pi$ from the light-cone sum rules (LCSR) using the pion light-cone distribution amplitudes. The accuracy of the underlying correlation function is upgraded by establishing the hard-collinear factorization formula at the leading power up to the next-to-leading order  in $\alpha_s$. Furthermore, the next-to-leading power contributions are incorporated at the leading order  by including the two-particle and three-particle higher-twist pion distribution amplitudes up to twist-4 accuracy.  We perform a comprehensive numerical analysis with a systematic evaluation of theoretical uncertainties, predicting $g_{D^{*}D^{*}\pi} = 5.71_{-0.89}^{+0.70} \text{ GeV}^{-1}$ and $g_{B^{*}B^{*}\pi} = 4.90_{-0.67}^{+0.59} \text{ GeV}^{-1}$. 
Finally, by parametrizing the $1/m_H$ (with $H=D,\,B$) power corrections to extract the universal static coupling $\hat{g} = 0.30 \pm 0.04$, we compare our results with previous theoretical determinations and experimental data, and discuss the current limitations in disentangling the heavy quark spin symmetry breaking effects.

\thispagestyle{empty}
\end{abstract}

\maketitle

\pagenumbering{arabic}

\section{Introduction}

A thorough understanding of low-energy strong interactions among heavy hadrons is essential for heavy-flavor phenomenology. In the heavy quark sector, the dynamics of hadrons containing a single charm  or bottom  quark are governed by the approximate Heavy Quark Spin Symmetry (HQSS)~\cite{Wise:1992hn}. This symmetry naturally predicts that the low-energy effective interactions between the heavy meson doublets and light pseudo-Nambu-Goldstone bosons are dictated by a universal strong coupling constant. Within the framework of Heavy Meson Chiral Perturbation Theory (HM$\chi$PT), the effective Lagrangian is built to describe the interactions between heavy vector mesons and light pseudoscalar mesons~\cite{Casalbuoni:1996pg,Yan:1992gz,Burdman:1992gh}. In the strict HQSS limit, the coupling constant $g_{H^*H^*\pi}$ for the vector-to-vector transition is degenerate with $g_{H^*H\pi}$ for the vector-to-pseudoscalar transition. However, the $1/m_Q$ corrections unambiguously break this spin symmetry, manifesting not only as the hyperfine mass splitting between the ground state and the vector state but also as a theoretical deviation between $g_{H^*H^*\pi}$ and $g_{H^*H\pi}$. While the latter coupling, $g_{D^*D\pi}$, can be directly extracted from the kinematically allowed decay width of $D^* \to D\pi$ experimentally measured by the CLEO and BaBar collaborations~\cite{CLEO:2001sxb, BaBar:2013thi,BaBar:2013zgp}, the $H^* \to H^*\pi$ transition is kinematically  forbidden on-shell due to the phase space constraint. Consequently, $g_{H^*H^*\pi}$ cannot be directly extracted from physical on-shell decays and must instead be determined through non-perturbative theoretical methods.

The precise evaluation of these couplings is of direct phenomenological interest. In the widespread hadronic molecular interpretation of the newly discovered $XYZ$ exotic states, the long-range binding mechanism is provided predominantly by the One-Pion Exchange Potential (OPEP)~\cite{Tornqvist:1993ng}, as extensively discussed in recent comprehensive reviews~\cite{Chen:2016qju,Guo:2017jvc,Brambilla:2019esw}. Constructing the OPEP for heavy vector-vector meson configurations intimately relies on the $H^*H^*\pi$ coupling strength. Therefore, a reliable determination of $g_{D^*D^*\pi}$ and $g_{B^*B^*\pi}$ is required for evaluating the binding energies of threshold exotic candidates, such as the theorized $D^*\bar{D}^*$ molecule $Z_c(4020)$~\cite{BESIII:2013ouc,BESIII:2013mhi} and its bottom analogue $Z_b(10650)$~\cite{Belle:2011aa}. Furthermore, these couplings  dictate the magnitude of chiral loop corrections to the self-energies and transitions of heavy mesons.

Beyond the strong interaction dynamics, this coupling also plays a crucial role in rare weak decays, as it enters the residue of the $H^*$ pole in the vector $H^* \rightarrow \pi$ transition form factor. Exploring the rare weak decays of $B^*$ and $D^*$ mesons represents a largely uncharted frontier in flavor physics. With the rapid advancements in the instrumentation and techniques of heavy-flavor experiments, there is great hope that the weak decays of heavy vector mesons will be observed at the LHCb and Belle II facilities in the near future~\cite{Abudinen:2022dnw,Belle-II:2018jsg}. Several theoretical studies have investigated the semileptonic decays of $H^*$ mesons~\cite{Chang:2018sud,Ray:2019gkv,Wang:2024cyi}.

Over the past decades, the $g_{D^*D^*\pi}$ coupling was estimated using three-point QCD sum rules  (QCDSR) by calculating off-shell form factors for both the $D^*$ and $\pi$ channels~\cite{Carvalho:2005et,Bracco:2011pg}.  In Ref.~\cite{Wang:2007zm}, the $g_{D^*D^*\pi}$ coupling was evaluated within the framework of light-cone sum rules (LCSR)~\cite{Balitsky:1989ry,Belyaev:1994zk}
 at the leading order (LO) in $\alpha_s$, leaving room for systematic improvements through next-to-leading order (NLO) corrections. In addition,  lattice QCD (LQCD) extracts the strong pionic couplings of heavy mesons by rigorously evaluating the hadronic matrix elements of the axial-vector current on the discretized spacetime lattice~\cite{Can:2012tx,Becirevic:2012pf,Detmold:2012ge,Bernardoni:2014kla,Flynn:2015xna}. These ab initio results provide benchmarks for the $g_{B^*B^*\pi}$ couplings through HQSS. 

Building upon these previous efforts, LCSR  provide a systematic framework to extract hadronic coupling constants~\cite{Khodjamirian:1999hb,Li:2020rcg,Wang:2020yvi,Khodjamirian:2020mlb,Jin:2024zyy,Jiang:2024equ}. In this work, we perform a rigorous and updated derivation of the $g_{H^*H^*\pi}$ couplings within the LCSR approach. To ensure high theoretical accuracy, we carry out the calculation beyond the leading approximation by considering both perturbative and higher-twist corrections. Specifically, we establish the hard-collinear factorization formula based on soft-collinear effective theory (SCET)~\cite{Wang:2017ijn,Cui:2023yuq,Khodjamirian:2023wol} at the leading power (LP) up to the NLO, which involves the computation of the one-loop hard matching coefficients and the appropriate ultraviolet renormalization for the relevant operators. Furthermore, we incorporate the next-to-leading power (NLP) contributions at LO by including the two-particle and three-particle higher-twist pion light-cone distribution amplitudes (DAs). By matching the QCD-level spectral representations with the hadronic dispersion relations, we present a  numerical analysis for both the charm and bottom sectors.

The remainder of this paper is organized as follows. In Section~\ref{sec-2}, we introduce the definition of the strong coupling and derive the hadronic dispersion relation. In Section~\ref{sec-3}, we present the hard-collinear factorization of the correlation function at the leading power, detailing both the LO partonic matrix elements and the NLO QCD corrections. In Section~\ref{sec-4}, we construct the corresponding LCSR at the leading power. The evaluation of the next-to-leading power contributions from higher-twist pion DAs is presented in Section~\ref{sec-5}. We perform the numerical analysis and discuss the resulting coupling constants in Section~\ref{sec-6}. Section~\ref{sec-7} is reserved for a summary of our main observations. Finally, the definitions and expressions for the pion QCD DAs are collected in Appendix~\ref{appendix-1}.

\section{Theory summary for the $H^*H^*\pi$ coupling}
\label{sec-2}

The effective Lagrangian describing the strong interactions between heavy vector mesons and light pseudoscalar mesons  can be written  as~\cite{Cheng:2004ru}
\begin{align}
	\mathcal{L}= -\frac{1}{2}g_{H^*H^*\pi} \epsilon_{\mu\nu\alpha\beta} H_i^{*\mu} \partial^\nu P^{ij}  (\overrightarrow{\partial}^\alpha -\overleftarrow{\partial}_\alpha)  H_j^{*\beta\dagger} \,,
\end{align}
where $H_\mu^*$ denotes the heavy vector meson isospin doublet, with $H_\mu^* =(D_\mu^{*0},\,D_\mu^{*+})$ for the charm sector and $(B_\mu^{*-},\,\bar{B}_\mu^{*0})$ for the bottom sector. We adopt the convention $\epsilon_{0123}=-1$ for the Levi-Civita tensor, and the pion matrix is given by
\begin{align}
	P= 
	\begin{pmatrix}
		\pi^0/\sqrt{2} & \pi^+ \\
		\pi^- & -\pi^0/\sqrt{2}
	\end{pmatrix}\,,
\end{align}
where, in the limit $m_Q\to \infty$, the constant $g_{H^*H^*\pi}$ coincides, up to a prefactor, with the low energy parameter $\hat{g}$ in HM$\chi$PT~\cite{Casalbuoni:1996pg}. Accordingly, the hadronic matrix elements are parameterized as
\begin{align}
	\langle H^{*}(q,\epsilon) \pi(p) |\mathcal{L} | H^{*}(p+q,\eta) \rangle = g_{H^*H^*\pi}\, \epsilon_{\mu\nu\alpha\beta}\, \epsilon^{*\mu} \eta^\nu p^\alpha q^\beta\,,
\end{align}
where the couplings of different charge states are related by isospin symmetry (ignoring tiny isospin-breaking effects), 
\begin{align}
	g_{D^*D^*\pi}&\equiv g_{D^{*+}D^{*0}\pi^+}= -\sqrt{2}g_{D^{*+}D^{*+}\pi^0} =\sqrt{2} g_{D^{*0}D^{*0}\pi^0}=-g_{D^{*0}D^{*+}\pi^-} \,,\nn
	g_{B^*B^*\pi}&\equiv g_{\bar{B}^{*0}B^{*-}\pi^+}= -\sqrt{2}g_{\bar{B}^{*0}\bar{B}^{*0}\pi^0} =\sqrt{2} g_{B^{*-}B^{*-}\pi^0}=-g_{B^{*-}\bar{B}^{*0}\pi^-} \,.
\end{align}

Furthermore, this strong coupling is intimately related to the $H^*\to \pi$ transition. The relevant hadronic matrix element for the vector form factor $V(q^2)$ is parameterized as
\begin{align}
	\langle \pi(p)|\bar{q}\gamma_\mu Q|H^*(p+q,\eta) \rangle =\frac{-2\,V(q^2)}{m_{H^*}+m_\pi} \epsilon_{\mu\rho\alpha \beta} \eta^\rho p^\alpha q^\beta\,.
\end{align}
Based on the hadronic dispersion relation, the strong coupling $g_{H^*H^*\pi}$ can be extracted from the residue of the $H^*$ pole via
\begin{align}
	g_{H^*H^*\pi}=\frac{2\,m_{H^*}}{f_{H^*}(m_{H^*}+m_\pi)} \lim_{q^2\to m_{H^*}^2} \Big[(1-q^2/m_{H^*}^2) V(q^2)\Big]\,.
\end{align}

Now we construct the vacuum-to-pion correlation function defined with two local interpolating currents for two heavy vector mesons
\begin{align}
	\Pi_{\mu\nu}(p,q)= i\int d^4x\, e^{iq\cdot x} \langle \pi^+(p)| T\{\bar{u}(x) \gamma_{\mu\perp} Q(x), \bar{Q}(0)\gamma_{\nu\perp}d(0) \} |0 \rangle\,, \label{cfun}
\end{align}
where
\begin{align}
	\gamma_{\mu\perp}=\gamma_\mu-\frac{\slashed{\bar{n}}}{2} n_\mu- \frac{\slashed{n}}{2} \bar{n}_\mu\,, \quad  p_\mu=\frac{\bar{n}\cdot p}{2} n_\mu\,,
\end{align}
with the light-cone vectors satisfying $n\cdot \bar{n}=2$ and $n^2=\bar{n}^2=0$. Taking advantage of the definition for the decay constant,
\begin{align}
	\langle 0| \bar{q} \gamma_{\mu\perp} Q| H^*(q,\epsilon) \rangle =f_{H^*} m_{H^*} \epsilon_{\mu\perp}\,,
\end{align}
and inserting the complete set of intermediate states with $H^*$ quantum numbers twice into the correlation function in Eq.~(\ref{cfun}), we obtain the hadronic dispersion relation
\begin{align}
	\Pi_{\mu\nu}(p,q)
	= \epsilon_{\mu\nu pq}\biggl\{\frac{g_{H^*H^*\pi} f_{H^*}^2 m_{H^*}^2}{[m_{H^*}^2-(p+q)^2] [m_{H^*}^2-q^2]}  +\iint_{\Sigma_h} ds_1ds_2 \frac{\rho^h(s_1,s_2)}{(s_1-(p+q)^2)(s_2-q^2)}\biggr\}\,,
	\label{hadronic}
\end{align}
where $\Sigma_h$ stands for the physical continuum region of higher resonances and continuum states, and  the  subtraction terms, which will vanish after performing the double Borel transformation, are dropped.

\section{The hard-collinear factorization at LP}
\label{sec-3}

\subsection{The hard-collinear factorization at LO}
To compute the correlation function in Eq.~(\ref{cfun}) at the partonic level using the light-cone operator product expansion (OPE), we treat the propagating heavy quark as highly virtual and consistently adopt the $\overline{\text{MS}}$ scheme for the heavy quark mass $m_Q$ throughout the calculation. Furthermore, we employ the following power counting scheme for the external momenta
\begin{align}
	\bar{n}\cdot p\sim \mathcal{O}(m_Q), \quad |(p+q)^2-m_Q^2| \sim |q^2-m_Q^2| \sim \mathcal{O}(m_Q^2)\,,
\end{align}
and evaluate the partonic matrix element
\begin{align}
	F_{\mu\nu}(p,q,u)= i\int d^4x\, e^{iq\cdot x} \langle q(up) \bar{q}(\bar{u}p)| T\{\bar{d}(x) \gamma_{\mu\perp} Q(x), \bar{Q}(0)\gamma_{\nu\perp}u(0) \} |0 \rangle \,.
\end{align}
Matching this amplitude onto the SCET at LO yields
\begin{align}
	F_{\mu\nu}^{(0)}(p,q,u)&= \frac{n\cdot q}{u(p+q)^2+\bar{u}q^2-m_Q^2} \,\bar{q}(up) \frac{\slashed{\bar{n}}}{2} \gamma_{\mu\perp} \gamma_{\nu\perp} q(\bar{u}p) \nn
	&=\sum_{i=1,E}H_i^{(0)}\big((p+q)^2,q^2,u'\big) \otimes \langle O_i\rangle^{(0)}(u, u') \,.
\end{align}
Here, the result is expressed as a convolution of the LO hard kernel $H_i^{(0)}$ with the tree-level on-shell matrix elements of the SCET operators $O_i$. The corresponding hard kernels are identical,
\begin{align}
	H_1^{(0)}=H_E^{(0)}=\frac{n\cdot q}{u'(p+q)^2+\bar{u}'q^2-m_Q^2}\,, 
	\label{tree-kernel}
\end{align}
and the physical operator $O_1$ and the evanescent operator $O_E$ are defined as
\begin{align}
	O_1(u')&= \frac{\bar{n}\cdot p}{2\pi}\int  d\tau \,e^{-iu' \tau \bar{n}\cdot p} \bar{\chi}(\tau \bar{n}) i\epsilon_{\perp\mu\nu} \frac{\slashed{\bar{n}}}{2} \gamma_5 \chi(0) \,,\nn
	O_E(u')&= \frac{\bar{n}\cdot p}{2\pi}\int  d\tau \,e^{-iu' \tau \bar{n}\cdot p} \bar{\chi}(\tau \bar{n}) \Big(\frac{[\gamma_{\mu\perp},\gamma_{\nu\perp}]}{2} +i\epsilon_{\perp\mu\nu}\Big) \frac{\slashed{\bar{n}}}{2} \gamma_5 \chi(0)\,,
\end{align}
with $\epsilon_{\perp\mu\nu}=\epsilon_{\mu\nu\bar{n}n}/2$. 
At tree level, the partonic matrix element of the physical operator yields the corresponding spinor structure multiplied by $\delta(u-u')$, effectively eliminating the convolution integral, while the evanescent operator contribution vanishes.
By employing the definition of the twist-two pion SCET DA 
\begin{align}
	\langle\pi(p)|\bar{\chi}(x_+)\slashed{\bar{n}}\gamma_5 \chi(0) |0\rangle =-if_\pi \,\bar{n}\cdot p \int_0^1du \,e^{iup\cdot x_+} \,\phi_2(u, \mu)\,,
\end{align}
with $x_+=n\cdot x\,\bar{n}/2$, we find the hard-collinear factorization formula for the correlation function (\ref{cfun}) at LO
\begin{align}
	\Pi_{\mu\nu}^{(0)}(p,q)=\frac{ f_\pi}{2}  \, \bar{n}\cdot p\, \epsilon_{\perp\mu\nu} \int_{0}^{1}du \, \phi_2(u,\mu)\, H_1^{(0)}\big((p+q)^2,q^2,u\big)\,.
\end{align}

\subsection{The hard-collinear factorization at NLO}

Including the $\mathcal{O}(\alpha_s)$ gluon radiative corrections to the correlation function (\ref{cfun}), the OPE result reads
\begin{align}
	\Pi_{\mu\nu}^{\rm LP}(p,q)=\frac{ f_\pi}{2}  \bar{n}\cdot p\, \epsilon_{\perp\mu\nu} \int_{0}^{1}du \,\phi_2(u,\mu) \Big[H_1^{(0)}\big((p+q)^2,q^2,u\big) +\frac{\alpha_sC_F}{4\pi} H_1^{(1)}\big((p+q)^2,q^2,u\big)  \Big] \,.
	\label{ope-nlo}
\end{align}
In Eq.~(\ref{ope-nlo}), the term $H_1^{(1)}$ represents the one-loop hard matching coefficient. To extract this coefficient, we perform a matching calculation between the full QCD and the SCET partonic amplitudes. Since we adopt dimensional regularization to regularize both ultraviolet (UV) and infrared (IR) divergences, the loop integrals for the bare matrix elements of the SCET operators are purely scaleless and evaluate to zero. Consequently, the IR divergences of the full QCD amplitude are precisely subtracted by the UV renormalization factors of the SCET operators, allowing the master formula for $H_1^{(1)}$ to be determined as
\begin{align}
	H_1^{(1)}=T_1^{(1)} -H_1^{(0)} Z_{11}^{(1)}-H_E^{(0)} Z_{E1}^{(1)}\,,
	\label{master}
\end{align}
where $T_1^{(1)}$ represents the bare one-loop  on-shell matrix element of the QCD operator; $Z_{ij}$ denotes the ultraviolet renormalization factors of the SCET operator. 

\begin{figure}
	\begin{center}
		\includegraphics[width=1.0 \columnwidth]{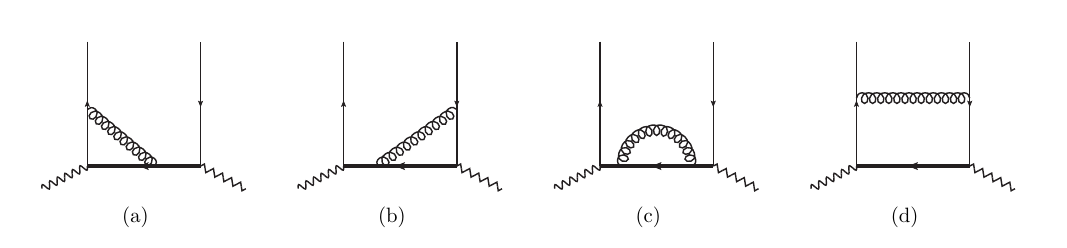} \\
		\caption{The one-loop Feynman diagrams for the twist-two pion DA contribution. }
		\label{twist-2-NLO}
	\end{center}
\end{figure}

We proceed to compute the NLO QCD correction to the four-point partonic amplitude $F_{\mu\nu}^{(1)}$ at LP displayed in Fig.~\ref{twist-2-NLO}. Specifically, Fig.~\ref{twist-2-NLO}(a) and \ref{twist-2-NLO}(b) represent the vertex corrections, Fig.~\ref{twist-2-NLO}(c) denotes the quark self-energy correction, and Fig.~\ref{twist-2-NLO}(d) shows the one-loop box diagram contribution.
The results are given by the following expressions, where we define the dimensionless kinematic variables $r_1 = (up+q)^2/m_Q^2$, $r_2 = q^2/m_Q^2$, and $r_3 = (p+q)^2/m_Q^2$,
\begin{align}
	F_{\mu\nu}^{(1a)}&=\frac{\alpha_sC_F}{4\pi} \Biggl\{ \Big[\frac{2(1-r_2)}{r_2-r_1} \big(L(r_1)-L(r_2)\big) -1 \Big] \Big(\frac{1}{\epsilon}+L\Big) 
	+\frac{2(1-r_2)}{r_1-r_2} \big[L_2(r_1)-L_2(r_2)  \big]  \nn
	& +\frac{(1-r_1)(1-r_1-2r_2)}{r_1(r_1-r_2)} L(r_1) - \frac{1-4r_2+3r_2^2}{(r_1-r_2) r_2} L(r_2) -4  \Biggr\} F_{\mu\nu}^{(0)}\,, \\
	F_{\mu\nu}^{(1b)}&=F_{\mu\nu}^{(1a)}(r_2\to r_3)\,, \\
	F_{\mu\nu}^{(1c)}
	&= - {\alpha_s \, C_F \over 4 \, \pi} \, \frac{1}{1-r_1}\bigg \{ \Big({1 \over \epsilon} +L\Big)(7 - r_1) -\frac{r_1^2-6r_1+1}{r_1^2}(1-r_1) L(r_1) \nn
	&- \frac{r_1^2-10r_1+1}{r_1} \bigg\}\, F_{\mu\nu}^{(0)} \,, \\
	F_{\mu\nu}^{(1d)}
	&= -{\alpha_s \, C_F \over 2 \, \pi} \, (1-r_1) F_{\mu\nu}^{(0)}\bigg \{ \Big[\frac{1-r_1}{(r_1-r_2)(r_1-r_3)}  L(r_1)  -\frac{1-r_2}{(r_1-r_2)(r_2-r_3)}  L(r_2)  \nn &+\frac{1-r_3}{(r_1-r_3)(r_2-r_3)}  L(r_3) \Big] \Big(\frac{1}{\epsilon} +L \Big) 
	-\frac{1-r_1}{(r_1-r_2)(r_1-r_3)}  \big[ L_2(r_1)  -5 L(r_1) \big] \nn
	&+\frac{1-r_2}{(r_1-r_2)(r_2-r_3)}  \big[L_2(r_2)  -5 L(r_2)\big]
	-\frac{1-r_3}{(r_2-r_3)(r_1-r_3)}  \big[L_2(r_3)  -5 L(r_3)\big] \bigg\}  \,,
\end{align}
where we introduce the auxiliary functions
\begin{align}
	L=\ln\frac{\mu^2}{m_Q^2}\,, \quad L(r_i)=\ln(1-r_i)\,, \quad L_2(r_i)=\ln^2(1-r_i)+{\rm Li}_2(r_i)\,.
\end{align}
Adding up different pieces together, we obtain $T_1^{(1)}$ in Eq.~(\ref{master}) as follows:
\begin{align}
	T_1^{(1)}&= \bigg\{2\bigg[ \frac{(r_1+1)(r_2+r_3)-r_1^2-2r_2r_3-1}{(r_1-r_2)(r_1-r_3)} L(r_1) 
	+\frac{1-r_1+r_2-r_3}{(r_1-r_2)(r_2-r_3)}(1-r_2) L(r_2)\nn &+\frac{1-r_1+r_3-r_2}{(r_1-r_3)(r_3-r_2)} (1-r_3)L(r_3) -\frac{3}{1-r_1} -\frac{3}{2} \bigg]  \Big(\frac{1}{\epsilon} +L \Big) \nn 
	&+\frac{2(1+r_1^2-r_3-r_1(r_2+r_3)+r_2(2r_3-1))}{(r_1-r_2)(r_1-r_3)}  L_2(r_1) \nn
	&-\frac{2(1-r_2)(1-r_1+r_2-r_3)}{(r_1-r_2)(r_2-r_3)}  L_2(r_2)
	-\frac{2(1-r_3)(1-r_1-r_2+r_3)}{(r_1-r_3)(r_3-r_2)} L_2(r_3) \nn
	&+\Big[\frac{1-r_1-2r_2}{r_1(r_1-r_2)} +\frac{2r_3+r_1-1}{r_1(r_3-r_1)} -\frac{10(1-r_1)}{(r_1-r_2)(r_1-r_3)} +\frac{r_1^2-6r_1+1}{(1-r_1)r_1^2}\Big] (1-r_1)L(r_1) \nn
	&-\Big[\frac{1-3r_2}{(r_1-r_2)r_2} -\frac{10(1-r_1)}{(r_1-r_2)(r_2-r_3)} \Big](1-r_2)L(r_2)\nn
	&-\Big[\frac{1-3r_3}{(r_1-r_3)r_3} -\frac{10(1-r_1)}{(r_1-r_3)(r_3-r_2)} \Big](1-r_3) L(r_3) +\frac{1-18r_1+9r_1^2}{r_1(1-r_1)}\bigg\} H_1^{(0)} \,,
\end{align}
where the LO hard kernel $H_1^{(0)}$ is given in Eq.~(\ref{tree-kernel}). The renormalization constant $Z_{E1}$ of the evanescent operator $O_E$ in Eq.~(\ref{master}) depends on the $\gamma_5$ scheme;  in  the naive dimensional regularization (NDR) scheme, $Z^{(1)}_{E1}$ is given in~\cite{Wang:2017ijn}
\begin{align}
	Z^{(1)}_{E1}(u,u')\big|_{\rm NDR}=\frac{\alpha_sC_F}{2\pi}(-4) \Big[ \frac{\bar{u}'}{\bar{u}} \theta(u'-u) +\frac{u'}{u}\theta(u-u')\Big]\,.
\end{align}
Then the contribution from the mixing of $O_E$ into $O_1$ yields
\begin{align}
	H_E^{(0)}\big((p+q)^2,q^2,u'\big)\otimes& Z^{(1)}_{E1}(u,u')\big|_{\rm NDR}=\frac{\alpha_sC_F}{2\pi}(-4)(1-r_1) \Big[ \frac{1-r_1}{(r_1-r_2)(r_1-r_3)} L(r_1) \nn
	&+ \frac{1-r_2}{(r_2-r_1)(r_2-r_3)} L(r_2) +\frac{1-r_3}{(r_3-r_1)(r_3-r_2)} L(r_3)  \Big] H_1^{(0)}\,.
\end{align}
It is then straightforward to find the one-loop hard kernel 
\begin{align}
	H_1^{(1),\rm NDR}&= \bigg\{2\bigg[ \frac{(r_1+1)(r_2+r_3)-r_1^2-2r_2r_3-1}{(r_1-r_2)(r_1-r_3)} L(r_1) 
	+\frac{1-r_1+r_2-r_3}{(r_1-r_2)(r_2-r_3)}(1-r_2) L(r_2)\nn &+\frac{1-r_1+r_3-r_2}{(r_1-r_3)(r_3-r_2)} (1-r_3)L(r_3) -\frac{3}{1-r_1} -\frac{3}{2} \bigg]  L \nn 
	&+\frac{2(1+r_1^2-r_3-r_1(r_2+r_3)+r_2(2r_3-1))}{(r_1-r_2)(r_1-r_3)}  L_2(r_1) \nn
	&-\frac{2(1-r_2)(1-r_1+r_2-r_3)}{(r_1-r_2)(r_2-r_3)}  L_2(r_2)
	-\frac{2(1-r_3)(1-r_1-r_2+r_3)}{(r_1-r_3)(r_3-r_2)} L_2(r_3) \nn
	&+\Big[\frac{1-r_1-2r_2}{r_1(r_1-r_2)} +\frac{2r_3+r_1-1}{r_1(r_3-r_1)} -\frac{2(1-r_1)}{(r_1-r_2)(r_1-r_3)} +\frac{r_1^2-6r_1+1}{(1-r_1)r_1^2}\Big] (1-r_1)L(r_1) \nn
	&-\Big[\frac{1-3r_2}{(r_1-r_2)r_2} -\frac{2(1-r_1)}{(r_1-r_2)(r_2-r_3)} \Big](1-r_2)L(r_2) \nn
	&-\Big[\frac{1-3r_3}{(r_1-r_3)r_3} -\frac{2(1-r_1)}{(r_1-r_3)(r_3-r_2)} \Big](1-r_3)L(r_3) +\frac{1-18r_1+9r_1^2}{r_1(1-r_1)}\bigg\} H_1^{(0)}\,.
	\label{h11-hard}
\end{align}
With the expression for the one-loop hard kernel at hand, it is crucial to verify the theoretical consistency of the factorization formula. To this end, we expand the twist-two pion DA $\phi_2$ in a series of Gegenbauer polynomials:
\begin{align}
	\phi_2(u,\mu)=6u\bar{u} \sum_{n=0}^{\infty} a_n(\mu) C_n^{3/2}(2u-1)\,,
\end{align}
where the scale dependence of the moments reads 
\begin{align}
	a_n(\mu)=E_{V,n}^{\rm NLO}(\mu,\mu_0) a_n(\mu_0) +\frac{\alpha_s(\mu)}{4\pi} \sum_{k=0}^{n-2} E_{V,n}^{\rm LO}(\mu,\mu_0) d_{V,n}^k(\mu,\mu_0) a_k(\mu_0)\,.
	\label{scale-depend}
\end{align}
Here, the NLO evolution implicitly depends on the chosen $\gamma_5$ scheme. The explicit expressions for the evolution functions $E_{V,n}^{\rm (N)LO}$, the mixing coefficients $d_{V,n}^k$, and the corresponding anomalous dimensions can be found in~\cite{Wang:2017ijn,Agaev:2010aq,Belitsky:2005qn}. It is worth noting that the $\gamma_5$ scheme dependence in Eq.~(\ref{h11-hard}) precisely cancels against that of the twist-2 pion DA. Therefore, $\Pi_{\mu\nu}^{\rm LP}$ in Eq.~(\ref{ope-nlo}) is  free of the $\gamma_5$ ambiguity.

The overall scale independence of the factorization formula can be  verified by making use of the evolution equation of the pion DA $\phi_2$
\begin{align}
	\frac{d}{d\ln\mu^2} \phi_2(u,\mu) =\int_0^1 du' \,V(u,u') \phi_2(u',\mu) \,,
	\label{scale-dependent}
\end{align}
where the LO ERBL kernel is given by~\cite{Lepage:1980fj,Efremov:1979qk}
\begin{align}
	V(u,u')=\frac{\alpha_s C_F}{2\pi} \Big[\frac{\bar{u}}{\bar{u}'} \Big(1+\frac{1}{u-u'} \Big)\theta(u-u') +\frac{u}{u'} \Big(1+\frac{1}{u'-u}\Big)\theta(u'-u)\Big]_+\,,
\end{align}
with the plus-prescription defined as
\begin{align}
	[f(u,u')]_+=f(u,u')-\delta(u-u') \int_0^1 dt\,f(t,u')\,.
\end{align}
Convoluting the LO hard kernel in Eq.~(\ref{tree-kernel}) with the evolution equation~(\ref{scale-dependent}) yields
\begin{align}
	&\int_0^1 du \,H_1^{(0)}\big((p+q)^2,q^2,u\big)  \frac{d}{d\ln\mu^2} \phi_2(u,\mu)\nn
	&= -\int_0^1du\,H^{(0)}\big((p+q)^2,q^2,u\big) \phi_2(u,\mu)  \frac{\alpha_s C_F}{2\pi} \bigg[ \frac{(r_1+1)(r_2+r_3)-r_1^2-2r_2r_3-1}{(r_1-r_2)(r_1-r_3)} L(r_1) \nn
	&+\frac{1-r_1+r_2-r_3}{(r_1-r_2)(r_2-r_3)}(1-r_2) L(r_2) +\frac{1-r_1+r_3-r_2}{(r_1-r_3)(r_3-r_2)} (1-r_3)L(r_3) -\frac{3}{2} \bigg] \,. 
\end{align}
From this result, it is evident that the factorization formula in Eq.~(\ref{ope-nlo}) is independent of the factorization scale $\mu$ at the one-loop level. That is to say,
\begin{align}
	\frac{d}{d\ln\mu^2}\Pi_{\mu\nu}^{\rm LP} =\mathcal{O}(\alpha_s^2) \,.
\end{align}

\section{The LCSR at LP}
\label{sec-4}

Now we derive the spectral representation of the factorization formula in Eq.~(\ref{ope-nlo})
\begin{align}
	\Pi_{\mu\nu}^{\rm LP}(p,q)=\frac{f_\pi}{2} \bar{n}\cdot p \,\epsilon_{\perp\mu\nu} \iint \frac{ds_1ds_2 }{[s_1-(p+q)^2](s_2-q^2)} \Big[\rho^{(0)}(s_1,s_2) +\frac{\alpha_sC_F}{4\pi}\rho^{(1)}(s_1,s_2)\Big] \,,
	\label{spectral-rep}
\end{align}
with the double spectral density defined as
\begin{align}
	\rho^{(0),(1)}(s_1,s_2)&=\frac{1}{\pi^2} {\rm Im}_{s_2}  {\rm Im}_{s_1} \int_0^1du \,\phi_2(u,\mu)\,H_1^{(0),(1)}(s_1,s_2,u)\,.
\end{align}
It is convenient to expand the pion DA as $\phi_2(u)=\sum c_ku^k$; then, the LO spectral density is given by \cite{Belyaev:1994zk}
\begin{align}
	\rho^{(0)}(s_1,s_2)=\frac{n\cdot q}{m_Q^2}\sum_{k=0}\frac{(-1)^{k+1}}{\Gamma(k+1)} c_k(\hat{s}_1-1)^k \delta^{(k)}(\hat{s}_1-\hat{s}_2)\theta(\hat{s}_1-1)\,,
	\label{lo-density}
\end{align}
with $\hat{s}_{1(2)}=s_{1(2)}/m_Q^2$, and $\delta^{(k)}(\hat{s}_1-\hat{s}_2)=d^k/d\hat{s}_1^k\, \delta(\hat{s}_1-\hat{s}_2)$. Since  the non-asymptotic effects in $\rho^{(1)}$ are much smaller than the overall theoretical uncertainty in the LCSR~\cite{Khodjamirian:1999hb}, we will use the asymptotic DA $\phi_2(u)=6u\bar{u}$ for the NLO spectral density. Taking advantage of the formulae for spectral representations in~\cite{Li:2020rcg}, we derive the analytical expression
\begin{align}
	&\rho^{(1)}(s_1,s_2)\nn&=3\frac{n\cdot q}{m_Q^2} \theta(\hat{s}_1-1) \biggl\{	
	\bigg[	(\hat{s}_1-1)(\hat{s}_2-1) \bigg(6\, {\rm Li}_2(1-\hat{s}_1) +2\, {\rm Li}_2(1-\hat{s}_2) +\frac{8}{3}\pi^2\nn
	&-2\big(\ln(\hat{s}_1-1)-\ln(\hat{s}_2-1)\big)^2 +4\ln \hat{s}_1 \ln(\hat{s}_1-1) -2\ln \hat{s}_1 \ln(\hat{s}_2-1) \nn
	&+\Big(\frac{1}{\hat{s}_1} -\frac{1}{\hat{s}_2} +2\ln \hat{s}_2\Big) \ln(\hat{s}_2-1) 
	-\Big(4+\frac{2}{\hat{s}_1}\Big)\ln(\hat{s}_1-1) 	\bigg) +4\hat{s}_2-5\hat{s}_1 \hat{s}_2\nn
	&+(\hat{s}_1+\hat{s}_2+4\hat{s}_1\hat{s}_2) \ln \hat{s}_1 
	+2(2-\hat{s}_1-\hat{s}_2) (3L+4) 
	+ \frac{\hat{s}_2}{\hat{s}_1}\bigg] \delta^{(2)}(\hat{s}_2-\hat{s}_1) \nn
	&+\theta(\hat{s}_2-1)(\hat{s}_1-1)(\hat{s}_2-1) \Big(4\ln\frac{\hat{s}_1-1}{\hat{s}_2-1}-2\ln\frac{\hat{s}_1}{\hat{s}_2}
	 +\frac{1}{\hat{s}_1}-\frac{1}{\hat{s}_2} \Big) \frac{d^3}{d\hat{s}_2^3} \ln|\hat{s}_2-\hat{s}_1| \biggr\}\,.
	 \label{lenthy}
\end{align}
By introducing the new dimensionless variables 
\begin{align}
	r=\frac{\hat{s}_2-1}{\hat{s}_1-1}\,, \quad \sigma=\hat{s}_1+\hat{s}_2-2\,,
\end{align}
we can transform the continuous logarithmic term algebraically by incorporating the Jacobian. Meanwhile, for the terms containing the $\delta$-function and its derivatives, we perform integration by parts with respect to $r$. Since the surface terms vanish at the kinematic boundaries, the derivatives acting on the $\delta$-function are safely eliminated. Consequently, at the integrand level, Eq.~(\ref{lenthy}) reduces to
\begin{align}
	&\rho^{(1)}(r,\sigma)\nn
	&=3\frac{n\cdot q}{m_Q^2}\frac{(r+1)^2}{\sigma} \biggl\{ (3L +4) \delta(\sigma) \delta(r-1) +\theta(\sigma)\delta(r-1) \Big[ -2{\rm Li}_2\Big(-\frac{\sigma}{2} \Big) \nn
	&-  \ln\frac{\sigma+2}{2} \ln\frac{\sigma}{2} 
	+\frac{24+32\sigma+10\sigma^2+\sigma^3}{2(\sigma+2)^3} \ln\frac{\sigma}{2} -\frac{1}{2}\ln\frac{\sigma+2}{2} +\frac{-4+4\sigma+\sigma^2}{4(\sigma+2)^2} -\frac{2\pi^2}{3}\Big] \nn
	&+\theta(\sigma) \theta(r) \frac{r}{r+1} \Big(-4\ln r-2\ln\frac{\sigma+r+1}{r\sigma+r+1} + \frac{r+1}{\sigma+r+1} -\frac{r+1}{r\sigma+r+1}\Big) \frac{d^3}{dr^3} \ln|r-1|\biggr\}\,.
	\label{nlo-density}
\end{align}

By matching the QCD factorization formula (\ref{spectral-rep}) with the hadronic representation (\ref{hadronic}) and performing the double Borel transformation with respect to the variables $(p+q)^2\to M_1^2$ and $q^2\to M_2^2$ (setting	$M_1^2=M_2^2=2M^2$), the resulting sum rule reads
\begin{align}
	g_{H^*H^*\pi}^{\rm LP} f_{H^*}^2 &=-\frac{f_\pi}{m_{H^*}^2} e^{\frac{m_{H^*}^2}{M^2}} \iint_{\Sigma} ds_1 ds_2\, e^{-\frac{s_1+s_2}{2M^2}}  \frac{1}{n\cdot q}\Big[\rho^{(0)}(s_1,s_2) +\frac{\alpha_sC_F}{4\pi}\rho^{(1)}(s_1,s_2)\Big] \nn
	&=-\frac{f_\pi}{m_{H^*}^2} e^{\frac{m_{H^*}^2}{M^2}}   \Big[\mathcal{F}_{\rm LP}^{(0)}(M^2,s_0) +\frac{\alpha_sC_F}{4\pi} \mathcal{F}_{\rm LP}^{(1)}(M^2,s_0)\Big]\,.
	\label{lp-result}
\end{align}
The general parameterization of the boundary $\Sigma$ is given by~\cite{Balitsky:1989ry}:
\begin{align}
	\Big(\frac{s_1}{s_*}\Big)^\alpha +\Big(\frac{s_2}{s_*}\Big)^\alpha \leq 1\,,
\end{align}
where $s_*$ is adjusted to provide equal diagonal intervals. The duality region possesses a smooth border crossing across the diagonal $s_1=s_2$. Thus, the integral of the LO spectral density $\rho^{(0)}$ in Eq.~(\ref{lo-density}) (and including the NLP spectral density in Eq.~(\ref{nlp-density})) is independent of the shape of the region. However, since $\rho^{(1)}$ contains nonvanishing terms (the last line in Eq.~(\ref{nlo-density})) in the off-diagonal region, the choice of the region's shape can lead to a difference. In the following, we choose the triangle region ($\alpha=1$, $s_*=2s_0$) as our default choice. Previous work~\cite{Neubert:1991sp,Blok:1992fc} has argued for the necessity of setting a triangular duality region in other frameworks of sum rules. Whether this conclusion carries over to our case is still unclear. We will further discuss the errors associated with the dual region selection in the subsequent numerical analysis.

The major advantage of choosing the triangle region is that it allows us to obtain the analytical results of the $r$-integral directly. We find
\begin{align}
	\mathcal{F}_{\rm LP}^{(0)}(M^2,s_0)&=-M^2\big(e^{-\frac{m_Q^2}{M^2}} -e^{-\frac{s_0}{M^2}} \big) \phi_2\Big(\frac{1}{2},\mu\Big) \,,\label{F0-fun}\\
	\mathcal{F}_{\rm LP}^{(1)}(M^2,s_0)&=\int_{2m_Q^2}^{2s_0} ds\, e^{-\frac{s}{2M^2}}\, \Big[ f_{\text{mass}}\Big(\frac{s}{m_Q^2}-2\Big) + f_{\text{cont}}\Big(\frac{s}{m_Q^2}-2\Big)  \Big]\,,
\end{align}
where
\begin{align}
	f_{\text{mass}}(\sigma) &= 3 \Big(3L +4\Big) \delta(\sigma-0^+)  \,, \nn
	f_{\text{cont}}(\sigma) &= 3\,\theta(\sigma) \Big[-2{\rm Li}_2\Big(-\frac{\sigma}{2} \Big)+{\rm Li}_2\Big(-\frac{\sigma}{\sigma+2}\Big)
	-{\rm Li}_2\Big(\frac{\sigma}{\sigma+2}\Big) \nn
	&- \ln\frac{\sigma+2}{2} \ln\frac{\sigma}{2} +\frac{\sigma^3+10\sigma^2+32\sigma+24}{2(\sigma+2)^3} \ln\frac{\sigma}{2} -\frac{1}{2}\ln\frac{\sigma+2}{2} \nn
	&-\frac{2(\sigma+1)(\sigma+4)}{(\sigma+2)^3}\ln(\sigma+1) + \frac{5\sigma^2+20\sigma+28}{4(\sigma+2)^2} -\frac{\pi^2}{6}\Big] \,. \label{f-fun}
\end{align}
The $f_{\text{mass}}(\sigma)$ term is the one-loop mass conversion factor between the $\overline{\text{MS}}$ and pole mass schemes, and the $f_{\text{cont}}(\sigma)$ term represents the genuine continuous dynamical one-loop correction.
If the purely asymptotic form of $\phi_2$ were adopted, the sum rule in Eq.~(\ref{lp-result}) would be exactly scale-independent at $\mathcal{O}(\alpha_s)$.
However, as will be specified in the numerical analysis, we adopt the more realistic non-asymptotic models for the LO contribution in Eq.~(\ref{F0-fun}), which inevitably leaves a slight residual scale dependence.

Alternatively, Ref.~\cite{Pullin:2021ebn} introduces a systematic technique to evaluate such double dispersion integrals by resolving second-type singularities. This approach expands the density into explicit poles and evaluates the reconstructed real and imaginary parts using Cauchy principal value integrals over the geometrically split duality regions. While this method is highly general, it relies heavily on numerical integrations rather than yielding a closed-form analytical expression like Eq.~(\ref{f-fun}).

\section{The LCSR for the higher-twist corrections}
\label{sec-5}

We now proceed to compute the higher-twist corrections to the coupling from the two-particle and three-particle pion DAs at LO, up to twist-four accuracy. This calculation relies on the light-cone expansion of the heavy-quark propagator  in the background field~\cite{Balitsky:1987bk}, which reads
\begin{eqnarray}
	&&\langle\,0\,|\,T\{Q(x),\,\bar{Q}(0)\}\,|\,0\,\rangle \supset \int\frac{d^4k}{(2\,\pi)^4}\,e^{-i\,k\cdot x}\,\frac{i\,(\slashed{k}+m_Q)}{k^2-m_Q^2} \nonumber\\
	&&+i\,g_s\int\frac{d^4k}{(2\pi)^4}\,e^{-i\,k\cdot x}\int_0^1du \,\Big[ {u \, x_{\mu} \over k^2-m_Q^2 } \, G^{\mu \nu}(u\, x) \, \gamma_{\nu}
	- {\slashed{k} + m_Q \over 2\,(k^2-m_Q^2)^2} \, G^{\mu \nu}(u\, x) \, \sigma_{\mu \nu}\Big]\,.
\end{eqnarray}
Substituting this expansion into the correlation function in Eq.~(\ref{cfun}) and employing the pion DA definitions collected in Appendix \ref{appendix-1}, we obtain the corresponding factorization formulas for the higher-twist contributions:
\begin{align}
	\Pi_{\mu\nu}^{\rm 2PHT}(p,q)
	&=-f_\pi \epsilon_{\mu\nu pq} \int_{0}^{1} du\,\biggl\{\frac{1}{\big((up+q)^2-m_Q^2\big)^2}\Big( \frac{1}{4} \phi_4(u)-\frac{1}{3}  m_Q \mu_\pi \phi_3^\sigma(u)\Big) \nn &-\frac{m_Q^2\phi_4(u)}{2\big((up+q)^2-m_Q^2\big)^3}\biggr\} \,, \label{2pht}\\
	\Pi_{\mu\nu}^{\rm 3PHT}(p,q)
	&=-f_\pi \epsilon_{\mu\nu pq}  \int_0^1dv \int[d\alpha_i] \frac{(1-2v)\Phi_4(\alpha_i)-\tilde{\Phi}_4(\alpha_i)}{\big((\alpha_vp+q)^2 -m_Q^2\big)^2} 	 \nn
	&=-f_\pi \epsilon_{\mu\nu pq} \int_{0}^{1} du\,\frac{1}{\big((up+q)^2-m_Q^2\big)^2} \,\bar{\Phi}_4(u)\,, \label{3pht}
\end{align}
where the integrated three-particle DA is defined as
\begin{align}
	\bar{\Phi}_4(u)=\int_0^ud\alpha_q \int_{u-\alpha_q}^{1-\alpha_q} \frac{d\alpha_g}{\alpha_g} \Big[\Big(1-2\frac{u-\alpha_q}{\alpha_g}\Big) \Phi_4(\alpha_i) -\tilde{\Phi}_4(\alpha_i)\Big]_{\alpha_{\bar{q}}= 1-\alpha_q-\alpha_g}\,.
\end{align}
Similar to Eq.~(\ref{lo-density}), we perform a Taylor expansion of the higher-twist DAs  $\psi(u)=\sum c_ku^k$ entering Eq.~(\ref{2pht}) and (\ref{3pht}). The general expression for the double spectral density reads \cite{Li:2020rcg}
\begin{align}
	&\frac{1}{\pi^2} {\rm Im}_{s_2}  {\rm Im}_{s_1}\int_0^1du \,\frac{\psi(u)}{(us_1+\bar{u}s_2-m_Q^2)^n}\nn
	&=\frac{1}{\Gamma(n)} \frac{d^{n-1}}{(dm_Q^2)^{n-1}}\sum_{k=0}\frac{(-1)^{k+1}}{\Gamma(k+1)} c_k(s_1-m_Q^2)^k \delta^{(k)}(s_1-s_2)\theta(s_1-m_Q^2)\,.
	\label{nlp-density}
\end{align}
We can then derive the LO LCSR for the higher-twist corrections
\begin{align}
	g_{H^*H^*\pi}^{\rm NLP} f_{H^*}^2 &=-\frac{f_\pi}{m_{H^*}^2} \, e^{\frac{m_{H^*}^2}{M^2}}  \, \Big[\mathcal{F}_{\rm NLP}^{\rm 2P}(M^2,s_0) + \mathcal{F}_{\rm NLP}^{\rm 3P}(M^2,s_0)\Big]\,,
	\label{nlp-result}
\end{align}
where
\begin{align}
	\mathcal{F}_{\rm NLP}^{\rm 2P}(M^2,s_0)&=e^{-\frac{m_Q^2}{M^2}}\, \Big[\frac{1}{4} \Big(1+\frac{m_Q^2}{M^2} \Big) \phi_4\Big(\frac{1}{2}\Big) -\frac{1}{3} m_Q\,\mu_\pi\, \phi_3^\sigma\Big(\frac{1}{2}\Big) \Big]\,, \\
	\mathcal{F}_{\rm NLP}^{\rm 3P}(M^2,s_0)&=e^{-\frac{m_Q^2}{M^2}}\, \bar{\Phi}_4\Big(\frac{1}{2}\Big)\,.
\end{align}

Putting Eq.~(\ref{lp-result}) and (\ref{nlp-result}) together, the final expression for the $H^*H^*\pi$ coupling can be written as
\begin{align}
	g_{H^*H^*\pi} f_{H^*}^2 
	=-\frac{f_\pi}{m_{H^*}^2} e^{\frac{m_{H^*}^2}{M^2}}   \Big[\mathcal{F}_{\rm LP}^{(0)}(M^2,s_0) +\frac{\alpha_sC_F}{4\pi} \mathcal{F}_{\rm LP}^{(1)}(M^2,s_0) +\mathcal{F}_{\rm NLP}^{\rm 2P}(M^2,s_0) + \mathcal{F}_{\rm NLP}^{\rm 3P}(M^2,s_0)\Big]\,.
	\label{final}
\end{align}

\section{Numerical analysis}
\label{sec-6}

In this section, we detail the input parameters required for the LCSR in Eq.~(\ref{final}). The numerical values and uncertainties of all relevant QCD and hadronic parameters are summarized in Table~\ref{tab-1}.
\begin{table}[h]
	\setlength{\tabcolsep}{8pt}
	\centering 
	\begin{tabular}{|ccc|ccc|} 
		\hline 
		\hline
		Parameter & Value & Ref. & Parameter & Value & Ref. \\
		\hline
		$f_{D^*}$ & $228.5\pm 7.7$ MeV & \cite{FLAG:2024oxs,Lubicz:2017asp}& $f_{B^*}$ & $182.0\pm 4.4$ MeV &\cite{Lubicz:2017asp,FLAG:2024oxs}\\
		\hline
		$\overline{m}_c(\overline{m}_c)$ & $1.273\pm0.005 $ GeV & \cite{ParticleDataGroup:2026aaa}& $\overline{m}_b(\overline{m}_b)$ & $4.186\pm0.006 $ GeV &\cite{ParticleDataGroup:2026aaa}\\
		\hline
		$a_2(2\,{\rm GeV})$ & $0.116_{-0.020}^{+0.019}$ & \cite{RQCD:2019osh} & &  &  \\
		$f_\pi$ & $130.2\pm0.8$ MeV &\cite{FLAG:2024oxs} & $\mu_\pi(2{\rm GeV})$ & $2.84\pm 0.04$ GeV &\cite{FLAG:2024oxs}\\
		$f_{3\pi}(1\,{\rm GeV})$ & $(4.5\pm1.5) \cdot 10^{-3}\, {\rm GeV}^2 $ & \cite{Ball:2006wn} & $\omega_{3\pi}(1\,{\rm GeV})$ & $-1.5\pm 0.7$  & \cite{Ball:2006wn} \\
		$\delta_{\pi}^2(1\,{\rm GeV})$ & $0.18\pm0.06\, {\rm GeV}^2 $ & \cite{Ball:2006wn} &
		$\epsilon_\pi(1\,{\rm GeV})$ & $0.5\pm 0.3$  &\cite{Ball:2006wn}  \\
		\hline
		$\mu(D^*)$ & $1.5_{-0.5}^{+1.5} $ GeV &\cite{Khodjamirian:2009ys} &$\mu(B^*)$ & $3.0_{-0.5}^{+1.5} $ GeV &\cite{Khodjamirian:2011ub} \\
		$M^2(D^*)$ & $4.5\pm 1.0 \, {\rm GeV}^2$ &\cite{Khodjamirian:2009ys} &$M^2(B^*)$ & $16.0\pm 4.0\, {\rm GeV}^2 $ &\cite{Khodjamirian:2011ub}\\
		$s_0(D^*)$ & $7.0\pm0.5\, \, {\rm GeV}^2$ &\cite{Khodjamirian:2009ys} &$s_0(B^*)$ & $37.5\pm 2.5\, {\rm GeV}^2 $ &\cite{Khodjamirian:2011ub}\\
		\hline 
		\hline
	\end{tabular} 
	\caption{Numerical input values of the theory parameters employed in the LCSR.}
	\label{tab-1}
\end{table}
To determine the heavy vector meson decay constants, we combine the LQCD results for the pseudoscalar decay constants~\cite{FLAG:2024oxs} with the vector-to-pseudoscalar ratios~\cite{Lubicz:2017asp}, following the procedure outlined in Ref.~\cite{Khodjamirian:2020mlb}. While an alternative approach is to substitute the decay constants with their respective two-point QCD sum rules to achieve a partial cancellation of parameter uncertainties, the precision of contemporary LQCD determinations has reached a level where direct substitution yields highly competitive and reliable results. Furthermore, this direct approach completely avoids the additional systematic uncertainties introduced by the auxiliary two-point sum rules (such as the dependence on extra Borel parameters and continuum thresholds). Consistent with our theoretical setup, we utilize the running heavy quark masses $\overline{m}_Q(\mu)$ evaluated in the $\overline{\text{MS}}$ scheme.

The key parameter of the LCSR  is the value of $ \phi_2(1/2,\mu)$ for the LO and LP contribution. We adopt the following two phenomenological models for the pion twist-2 DA~\cite{RQCD:2019osh}:
\begin{align}
	\phi_2^{\rm(I)}(u)&=6u\bar{u}\big[1+a_2\,C_2^{3/2}(2u-1)\big] \,, \nn
	\phi_2^{\rm(II)}(u)&= \frac{\Gamma(2+2\beta)}{[\Gamma(1+\beta)]^2}\, u^\beta \bar{u}^\beta\,.
\end{align}
The first model is the expansion in Gegenbauer polynomials truncated after $n = 2$, where the second moment $a_2$  is adopted from the LQCD determination by the RQCD Collaboration~\cite{RQCD:2019osh}. The second model is based on a simple power-law parametrization, and its parameter $\beta$ is fixed by equating the second Gegenbauer moment to the value of $a_2$ in Model I, that is, $\beta=(7-18a_2)/(7+12a_2)$. The expressions for the higher-twist pion DAs in the conformal expansion are provided in  Appendix~\ref{appendix-1}. The parameter $\mu_{\pi}$ is determined using the light-quark mass combination $m_u + m_d$ given in~\cite{FLAG:2024oxs}. Following Eq.~(\ref{scale-depend}), the renormalization scale dependence of the twist-two parameters is evaluated at next-to-leading-logarithmic (NLL) accuracy. Additionally, the renormalization scale dependence of the higher-twist parameters at leading-logarithmic (LL) accuracy can be found in~\cite{Duplancic:2008ix}.
We vary the factorization scale in the range of $[1.0,\,3.0]$ GeV around a central value of $1.5$ GeV for $D^*D^*\pi$ coupling, and within $[2.5,\,4.5]$ GeV around a central value of $3.0$ GeV for $B^*B^*\pi$ coupling.
The values for the Borel parameters $M^2$ and the effective thresholds $s_0$ are adopted from the pseudoscalar channel sum rules~\cite{Khodjamirian:2009ys,Khodjamirian:2011ub}, which are consistent with those used in~\cite{Khodjamirian:2020mlb}. It is worth noting that for the vector channels considered in this work, the continuum thresholds are expected to undergo a kinematic shift of roughly $m_{H^*}^2 - m_H^2 \sim 0.5 \, \text{GeV}^2$. This expected shift is well accommodated within our quoted uncertainties for $s_0$ ($\pm 0.5 \, \text{GeV}^2$ for the $D^*$ channel and $\pm 2.5 \, \text{GeV}^2$ for the $B^*$ channel), thereby ensuring the reliability of the extracted sum rules.

\begin{table}[h]
	\setlength{\tabcolsep}{6pt}
	\centering 
	\begin{tabular}{|c|c|c|c|c|c|c|} 
		\hline 
		LCSR results &  Tw-2 LO  &   NLO (mass) &  NLO (cont.) & Tw-3 LO & Tw-4 LO  & Total \\
		\hline
		$f^2_{D^*}\,g_{D^*D^*\pi}$  & \makecell{$0.223$ (Model I) \\ $0.239$ (Model II)} & $-0.049$ & $0.033$ & $0.089$ & $-0.014$  & \makecell{0.282\\0.298}\\
		\hline
		$f^2_{B^*}\,g_{B^*B^*\pi}$  & \makecell{$0.104$ (Model I) \\ $0.109$ (Model II)}  & $-0.020$ & $0.029$ & $0.046$ & $-0.003$  & \makecell{$0.157$ \\ $0.162$ }\\
		\hline
	\end{tabular} 
	\caption{Numerical results for the products $f_{H^*}^2 g_{H^*H^*\pi}$ (in GeV) at the central input values.  The twist-2 NLO contribution is separated into the mass-scheme conversion piece [NLO (mass)] and the genuine continuous perturbative piece [NLO (cont.)].}
	\label{tab-2}
\end{table}
The numerical breakdown for the products of the strong couplings and the corresponding decay constants is presented in
Table~\ref{tab-2}.  As expected, the twist-2 LO contribution is dominant. 
The overall NLO corrections are found to be relatively small (around $5\%$ of the total) and exhibit opposite signs for the $D^*$ and $B^*$ channels. To understand this behavior, Table~\ref{tab-2} reveals that the mass-scheme conversion effect provides a sizable negative contribution that largely cancels against the positive genuine continuous NLO piece. Consequently, the sign flip between the $D^*$ and $B^*$ channels simply reflects the shifting of this cancellation point as the running mass $m_Q(\mu)$ and the associated scale $L$ evolve from the charm to the bottom sector, rather than an indicator of perturbative instability.
The twist-3 LO correction contributes approximately $40\%$ of the twist-2 LO value. Meanwhile, the twist-4 LO corrections remain extremely small, which ensures the convergence of the OPE within this framework. Furthermore, the choice between Model I and Model II introduces only a slight variation in the final totals, indicating that the higher Gegenbauer moments ($a_{n\ge4}$) are negligible. Therefore, we adopt Model II for our final predictions. 

\begin{figure}
	\begin{center}
		\includegraphics[width=1.0 \columnwidth]{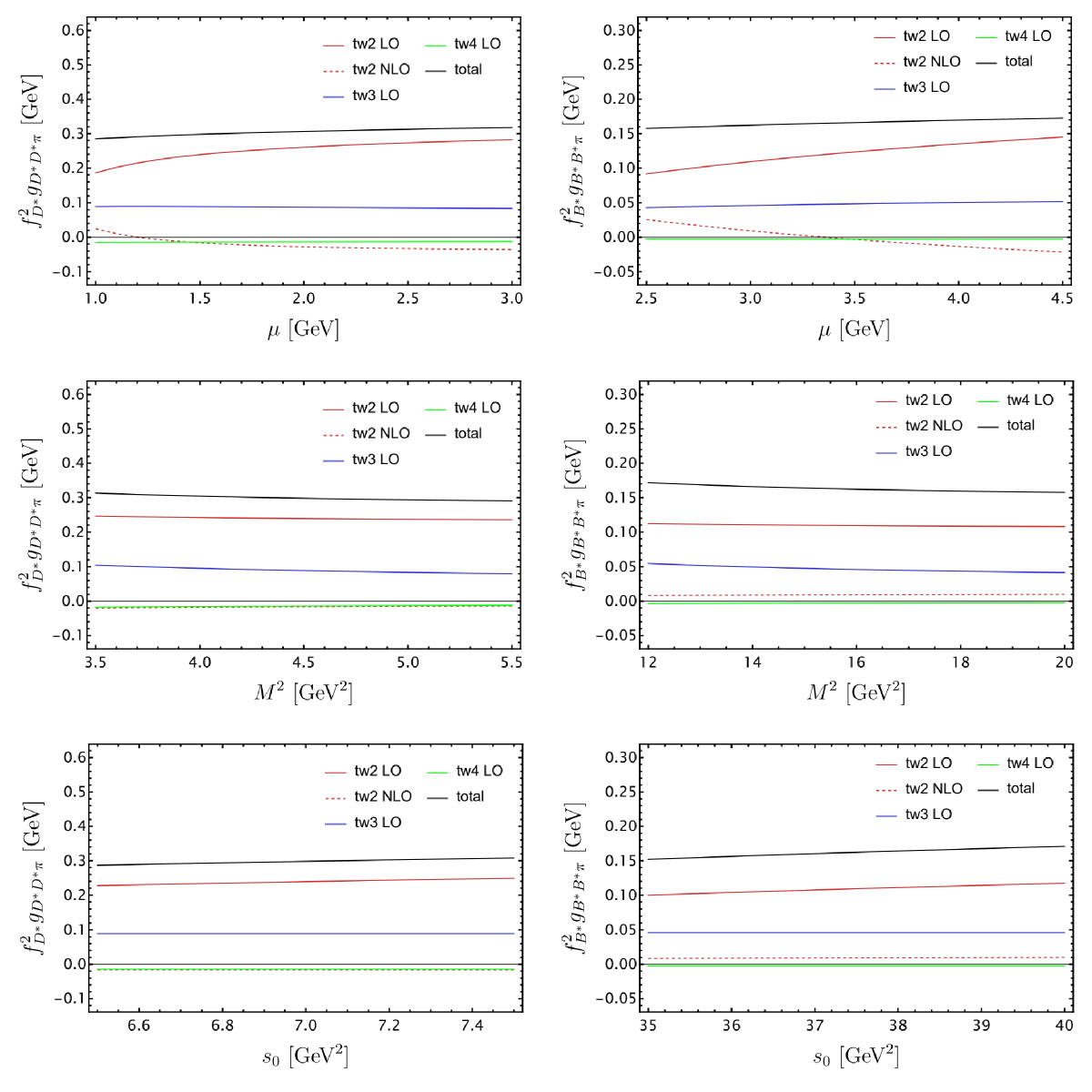} \\
		\caption{Dependence of the LCSR predictions for $f_{D^{*}}^{2}g_{D^{*}D^{*}\pi}$ (left column) and $f_{B^{*}}^{2}g_{B^{*}B^{*}\pi}$ (right column) on the factorization scale $\mu$, Borel mass $M^2$, and effective continuum threshold $s_0$, with all other input parameters fixed at their central values. }
		\label{depedence-line}
	\end{center}
\end{figure}
We now investigate the stability of $f_{H^*}^2 g_{H^*H^*\pi}$ with respect to variations in the scale $\mu$, the Borel mass $M^2$, and the threshold $s_0$, as shown in Fig.~\ref{depedence-line}. In each panel, the total prediction (black solid line) is decomposed into its individual components. The slight residual dependence on the factorization scale $\mu$ observed in the top panels is primarily attributable to two theoretical truncations: the omission of the twist-3 NLO corrections, and the fact that the twist-2 NLO calculation only accounts for the asymptotic distribution amplitude. For the individual NLO components, however, the scale dependence is pronounced, causing them to cross zero and change sign.  This pronounced scale dependence is a direct mathematical consequence of the logarithmic factor $L$ within $f_{\text{mass}}(\sigma)$ in Eq.~(\ref{f-fun}).
As demonstrated  in the middle and bottom panels, the total results exhibit stable Borel plateaus within the established working windows for $M^2$ and $s_0$.

By dividing out the heavy-meson decay constants, we extract the strong couplings:
\begin{align}
	g_{D^*D^*\pi}=5.71^{+0.70}_{-0.89}\,\, {\rm GeV}^{-1}\,, \quad 	g_{B^*B^*\pi}=4.90^{+0.59}_{-0.67}\, {\rm GeV}^{-1}\,.
	\label{results}
\end{align}

Table~\ref{tab-error} presents a comprehensive error budget for the extracted strong couplings.  The theoretical uncertainties in our framework arise from parametric inputs and systematic modeling approximations. To address the systematic uncertainties, we have incorporated several additional sources into this budget. First, while calculating the complete NLO corrections for the twist-3 DAs is non-trivial and beyond our current scope, we estimate its magnitude based on the $\mathcal{O}(\alpha_s)$ convergence rate at the twist-2 level. A conservative uncertainty of $10\%$ relative to the twist-3 LO contribution is thus assigned as $\Delta_{\text{Tw3-NLO}}$. Second, we estimate the theoretical uncertainty arising from the heavy quark mass scheme, denoted as $\Delta_{\text{mass}}$, by comparing the total couplings evaluated in the $\overline{\text{MS}}$ and pole mass schemes. The latter results can be obtained by setting $f_{\text{mass}}(\sigma) \to 0$ and $\overline{m}_Q\to m_Q^{\text{pole}}$ with $m_c^{\text{pole}} =1.67 \text{ GeV}$ and $m_b^{\text{pole}}= 4.78 \text{ GeV}$~\cite{ParticleDataGroup:2026aaa}. Third, the variations induced by the choice of the pion DA models (Model I vs. Model II) are  included as $\Delta_{\text{model}}=g_{\text{model I}}-g_{\text{model II}}$. Finally, to assess the systematic uncertainty introduced by the quark-hadron duality ansatz, we investigate the dependence of our results on the shape of the two-dimensional duality region~\cite{Khodjamirian:2020mlb,Pullin:2021ebn}. By evaluating the sum rules under three different boundary parameterizations (concave with $\alpha=1/2$, $s_*=4s_0$; triangular with $\alpha=1$, $s_*=2s_0$; and convex with $\alpha=2$, $s_*=\sqrt{2}s_0$), we find that the resulting deviation, which primarily originates from the sensitivity of the twist-2 NLO corrections to the integration boundaries, remains well under control and is safely incorporated as $\Delta \alpha$.

As revealed by the detailed breakdown in Table~\ref{tab-error},  alongside the heavy vector meson decay constants and the factorization scale $\mu$,  the choice of the heavy quark mass scheme ($\Delta_{\text{mass}}$) emerges as a dominant source of uncertainty. The pronounced sensitivity to the mass scheme and the factorization scale is an anticipated theoretical consequence of truncating the perturbative series at $\mathcal{O}(\alpha_s)$. In contrast, the variations induced by the Borel parameter $M^2$ and the effective continuum threshold $s_0$ are relatively moderate, corroborating the stability of the sum rules within the chosen working windows.

\begin{table}[h]
	\centering 
	\resizebox{\textwidth}{!}{%
		\begin{tabular}{|c|c|c|c|c|c|c|c|c|c|c|c|} 
			\hline 
			Coupling & Central & $\Delta f_{H^*}$ & $\Delta \mu$ &  $\Delta M^2$&  $\Delta s_0$ & $\Delta a_2$ & $\Delta_{\text{Tw3-NLO}}$ & $\Delta_{\text{mass}}$ &  $\Delta_{\text{model}}$  & $\Delta \alpha$  & Total \\
			\hline
			$g_{D^*D^*\pi}$ & $5.71$ & $_{-0.37}^{+0.41}$ & $_{-0.24}^{+0.38}$& $_{-0.15}^{+0.29}$ & $_{-0.22}^{+0.19}$  & $_{-0.10}^{+0.11}$ & $\pm 0.17$ & $_{-0.52}^{+0.00}$ & $_{-0.31}^{+0.00}$ & $_{-0.33}^{+0.11}$  & $_{-0.89}^{+0.70}$  \\
			\hline
			$g_{B^*B^*\pi}$  &$4.90$& $_{-0.23}^{+0.25}$  & $_{-0.14}^{+0.32}$  &$_{-0.14}^{+0.29}$ & $_{-0.31}^{+0.26}$ & $_{-0.06}^{+0.07}$ & $\pm 0.14$ &$_{-0.44}^{+0.00}$ & $_{-0.15}^{+0.00}$ & $_{-0.15}^{+0.09}$   & $_{-0.67}^{+0.59}$ \\
			\hline
		\end{tabular} 
	}
	\caption{Detailed breakdown of the theoretical uncertainties for the extracted strong couplings $g_{D^{*}D^{*}\pi}$ and $g_{B^{*}B^{*}\pi}$ (in  $\text{GeV}^{-1}$). The total uncertainties are obtained by adding all individual variations in quadrature, including the input parameters, the truncated NLO twist-3 corrections ($\Delta_{\text{Tw3-NLO}}$), the mass scheme $\Delta_{\text{mass}}$, DA model choice ($\Delta_{\text{model}}$), and systematic errors from the duality region shape ($\Delta \alpha$).}
	\label{tab-error}
\end{table}

As previously mentioned, the strong coupling $g_{H^*H^*\pi}$ is intimately related to the universal low-energy constant $\hat{g}$ in HM$\chi$PT, which dictates the interactions between heavy mesons and pseudo-Nambu-Goldstone bosons in the chiral and static limits. 
In the  heavy quark limit, this coupling satisfies the relation $g_{H^*H^*\pi} = 2\hat{g}/f_\pi$. To demonstrate the theoretical consistency of our framework, it is instructive to examine the asymptotic behavior of the derived LCSR. By applying the standard heavy-quark scaling relations to the LO amplitude,
\begin{align}
	m_{H^*} = m_Q + \bar{\Lambda}\,, \quad M^2 = 2m_Q\tau\,, \quad s_0 = m_Q^2 + 2m_Q\omega_0\,, \quad f_{H^*} = \frac{\hat{f}}{\sqrt{m_Q}}\,,
\end{align}
where $\bar{\Lambda}$ and $\hat{f}$ are the binding energy and the static decay constant of the heavy meson in Heavy Quark Effective Theory (HQET), respectively, while $\tau$ and $\omega_0$ denote the heavy-mass-independent Borel parameter and continuum threshold. Retaining the leading power in $1/m_Q$, the sum rule in Eq.~(\ref{final}) analytically reduces to
\begin{align}
	g_{H^*H^*\pi} \simeq \frac{2}{f_\pi} \left[ \frac{f_\pi^2}{\hat{f}^2} e^{\bar{\Lambda}/\tau} \left\{ \tau\Big(1 - e^{-\omega_0/\tau}\Big)\phi_2\Big(\frac{1}{2}\Big) + \frac{\mu_\pi}{6}\phi_3^\sigma\Big(\frac{1}{2}\Big) - \frac{\phi_4(1/2)}{16\tau} \right\} \right]\,.
	\label{static-limit}
\end{align}
The expression inside the square brackets is exactly the static coupling $\hat{g}$ evaluated in HQET. Remarkably, the dynamical structure of this sum rule is completely identical to that of the $H^*H\pi$ coupling derived in the same limit (see Eq.~(5.10) in Ref.~\cite{Khodjamirian:2020mlb}), thereby  verifying the HQSS. 

However, performing this analytical expansion at NLO is highly nontrivial, as it requires resumming the large logarithms $\ln(\mu^2/m_Q^2)$. Therefore, following the standard practice in LCSR, we proceed to extract the static coupling via a numerical fit utilizing the full finite-mass sum rules. By combining our results with the vector-to-pseudoscalar transition couplings obtained in Ref.~\cite{Khodjamirian:2020mlb} ($g_{D^*D\pi}=14.1_{-1.2}^{+1.3}$ and $g_{B^*B\pi}=30.0_{-2.4}^{+2.6}$), we parameterize the $1/m_H$ power corrections to formulate the coupled equations~\cite{Boyd:1994pa}:
\begin{align}
	g_{H^*H\pi}=\frac{2m_H\hat{g}}{f_\pi}\Big(1+\frac{\delta_1}{m_H}\Big) \,, \quad 
	g_{H^*H^*\pi}=\frac{2\hat{g}}{f_\pi}\Big(1+\frac{\delta_2}{m_H}\Big) \,.
	\label{extract}
\end{align}
Fitting these relations yields:
\begin{align}
	\hat{g}=0.30\pm0.04\,, \quad \delta_1=1.22\pm 0.59 \, {\rm GeV} \,, \quad \delta_2=0.44\pm 0.50 \, {\rm GeV} \,,
\end{align}
with the correlation matrix for the parameters $(\hat{g}, \delta_1, \delta_2)$ given by
\begin{align}
	\begin{pmatrix}
	1.0 & -0.91 & -0.81 \\
	-0.91 & 1.0 & 0.74 \\
	-0.81 & 0.74 &1.0
\end{pmatrix}\,.
\label{para-correla}
\end{align}
The extracted value of $\hat{g}$ is highly consistent with previous LCSR studies~\cite{Belyaev:1994zk,Khodjamirian:2020mlb}. However, the $1/m_Q$ correction parameters $\delta_1$ and $\delta_2$, which encode the HQSS breaking effects, exhibit notably large uncertainties. As revealed by the correlation matrix in Eq.~(\ref{para-correla}), these large errors are primarily driven by the strong statistical anti-correlations between the static coupling $\hat{g}$ and the mass-correction parameters $\delta_i$. It should be noted that the four LCSR inputs are treated as independent in this fit. Since these theoretical values share common parametric dependencies, rigorously accounting for their correlations would modify the resulting error contours. In light of these large uncertainties, with $\delta_2$ remaining consistent with zero at the $1\sigma$ level, a conclusive assessment of the HQSS breaking effects is premature. This indicates that while the LCSR approach provides a reliable constraint on the overall couplings, a much higher theoretical precision or additional lattice constraints are required to stringently disentangle these higher-order power corrections from the static limit.

\begin{table}[h]
	\setlength{\tabcolsep}{8pt}
	\centering 
	\begin{tabular}{|c|c|c|c|} 
		\hline 
		Method  & $g_{D^*D^*\pi}\, [{\rm GeV}^{-1}]$ & $g_{B^*B^*\pi} \, [{\rm GeV}^{-1}]$ & $\hat{g}$ \\
		\hline
		LCSR \cite{Khodjamirian:2020mlb} &$7.27_{-0.62}^{+0.67}$ & $5.66_{-0.45}^{+0.49}$ & $0.30\pm0.02$\\
		\hline
		QCDSR \cite{Bracco:2011pg} & $4.3\pm0.5$ & - & - \\
		\hline
		LQCD, $N_f=2+1$ \cite{Detmold:2012ge} &-&-& $0.449\pm0.051$ \\
		\hline
		LQCD, $N_f=2$ \cite{Bernardoni:2014kla}&-&-& $0.492\pm0.029$ \\
		\hline
		LQCD, $N_f=2+1$ \cite{Flynn:2015xna}&-&-& $0.56\pm0.03\pm0.07$ \\
		\hline
		experiment  \cite{BaBar:2013thi} &-&-& $0.570\pm0.004\pm0.005$ \\
		\hline
		this work & $5.71_{-0.89}^{+0.70}$ & $4.90_{-0.67}^{+0.59}$ &$0.30\pm0.04$\\ 
		\hline
	\end{tabular} 
	\caption{Comparison of the extracted strong couplings $g_{D^*D^*\pi}$, $g_{B^*B^*\pi}$ (in $\text{GeV}^{-1}$) and the universal static coupling $\hat{g}$ with previous theoretical determinations and experimental data.}
	\label{tab-final}
\end{table}
A comprehensive comparison of our predictions with existing literature and experimental measurements is summarized in Table~\ref{tab-final}. It is worth noting that the previous LCSR results in Ref.~\cite{Khodjamirian:2020mlb} were obtained using the relation $g_{H^*H^*\pi}=g_{H^*H\pi}/\sqrt{m_{H^*}m_{H}}$ in the strict chiral and heavy quark limits.
The numerical deviations between our results and those in Ref.~\cite{Khodjamirian:2020mlb} primarily arise from two sources. First, the omission of twist-3 NLO corrections in our current treatment contributes to the difference. Second, the effects of HQSS breaking play a significant role; this interpretation is robustly supported by the fact that the relative deviation in the heavier $B^*$ channel is noticeably smaller than that in the $D^*$ channel, as expected for $1/m_Q$ corrections. Furthermore, our results are in rough agreement with the previous QCDSR calculation in Ref.~\cite{Bracco:2011pg}.

Regarding the universal static coupling $\hat{g}$, the experimental value is extracted from the natural linewidth of the $D^{*+} \rightarrow D^0\pi^+$ transition measured by the BaBar collaboration~\cite{BaBar:2013thi}. As shown in Table~\ref{tab-final}, both the LQCD and experimental values for $\hat{g}$ are  larger than our LCSR result.  For experimental extractions that rely on the leading-order HQSS relation, the positive $1/m_Q$ power corrections are naturally absorbed into the effective static coupling, thereby yielding a larger $\hat{g}$. 
To further investigate the origin of the tension with the static-limit LQCD results~\cite{Detmold:2012ge,Bernardoni:2014kla}, where the $1/m_Q$ corrections are absent by construction, we performed a standalone fit utilizing only our newly evaluated $g_{D^*D^*\pi}$ and $g_{B^*B^*\pi}$ couplings via the second formula in Eq.~(\ref{extract}). This independent fit yields $\hat{g} = 0.29 \pm 0.07$, which is highly consistent with the joint fit result. This internal consistency clearly indicates that the lower value of $\hat{g}$ is an intrinsic prediction of the LCSR framework for these vertices. The systematic deviation between the LCSR predictions and the static lattice evaluations likely points to inherent methodological differences, such as the approximations within the quark-hadron duality ansatz.

\section{Conclusion}
\label{sec-7}

In this work, we have presented an updated and systematic derivation of the strong couplings $g_{D^*D^*\pi}$ and $g_{B^*B^*\pi}$ utilizing the LCSR framework. To ensure a high degree of theoretical rigor, our calculation extends beyond the LO approximation. Specifically, we established the hard-collinear factorization formula at LP up to NLO in $\alpha_s$, and incorporated the NLP contributions from the two- and three-particle higher-twist pion DAs up to twist-4 accuracy.

Our numerical analysis reveals a well-behaved convergence of the OPE within this framework. The dominant contribution arises from the twist-2 LO terms, while the twist-3 LO corrections provide a sizable enhancement. The negligible size of the twist-4 LO effects further validates the truncation of the higher-twist series. Furthermore, the inclusion of NLO perturbative corrections, which amount to roughly  $5\%$ of the total amplitude and exhibit opposite signs in the charm and bottom sectors, proves vital for precision.  As revealed by our analysis, this distinct sign flip originates from a numerical cancellation between the mass-scheme conversion piece and the continuous perturbative contribution within the NLO spectral density. We have also comprehensively evaluated various sources of systematic uncertainties, including the choice of the heavy quark mass scheme, the truncation of higher-twist corrections, and the geometric shape of the duality boundary, confirming that they remain well under control and are accounted for in the overall error budget.

Ultimately, we extract the strong couplings to be $g_{D^*D^*\pi} = 5.71_{-0.89}^{+0.70}\,\text{GeV}^{-1}$ and $g_{B^*B^*\pi} = 4.90_{-0.67}^{+0.59}\,\text{GeV}^{-1}$. Furthermore, we explored the potential impact of HQSS breaking. By parameterizing the finite mass effects with $1/m_Q$ power corrections, we extracted the universal static coupling to be $\hat{g} = 0.30 \pm 0.04$. Notably, this value is lower than direct extractions from LQCD and experimental linewidth measurements.  While separating the $1/m_Q$ corrections partially explains the deviation from experiment, a standalone fit confirms that this lower value is fundamentally an intrinsic LCSR prediction, reflecting inherent methodological differences from static-limit lattice simulations, such as OPE truncations and duality approximations.

The precise determination of these dynamical couplings provides valuable input for heavy-flavor phenomenology. They are required to construct the OPEP between heavy vector mesons, directly impacting the evaluation of binding energies for threshold exotic candidates like the $Z_c(4020)$ and $Z_b(10650)$. Moreover, as these couplings govern the residue of the $H^*$ pole in vector-to-pion transition form factors, they are necessary for studying rare weak decays of $B^*$ and $D^*$ mesons anticipated at LHCb and Belle II. Finally, while the LCSR approach offers reliable constraints on the overall couplings, the parameters dictating the HQSS breaking effects still suffer from substantial uncertainties due to statistical anti-correlations. Achieving higher theoretical precision will require incorporating the NLO perturbative corrections to higher-twist DAs. Combining such theoretical advancements with emerging LQCD data will be important to stringently isolate these higher-order power corrections from the static limit.

\section*{Acknowledgments}
This work was supported in part by the National Natural Science Foundation of China under Grant No. 12105112.

\appendix

\section{Light-Cone Distribution Amplitudes of the Pion}
\label{appendix-1}

The definitions of the two-particle and three-particle pion DAs up to twist-4 accuracy are given as follows~\cite{Ball:2006wn}:
\begin{align}
	&\langle\,\pi(p)\,|\,\bar{q}(x)\,\gamma_\mu\,\gamma_5\,q(0)\,|\,0\,\rangle \nn
	&=-i\,f_\pi\int_0^1 du \,e^{i\,u\,p\cdot x}\,\Big[ \Big(\phi_2(u)+\frac{x^2}{16}\,\phi_4(u)\Big)p_\mu + \frac{x_\mu}{2 p\cdot x} \psi_4(u)\Big]\,, \\
	&\langle\,\pi(p)\,|\,\bar{q}(x)\,\gamma_5\,q(0)\,|\,0\,\rangle
	=-if_\pi\,\mu_\pi\, \int_0^1 du \,e^{i\,u\,p\cdot x}\,\phi_3^p(u)\,,\\
	&\langle\,\pi(p)\,|\,\bar{q}(x)\,\sigma_{\mu\nu}\,\gamma_5\,q(0)\,|\,0\,\rangle
	=\frac{i}{6}\,f_\pi\, \mu_\pi\,(p_\mu\,x_\nu-p_\nu\,x_\mu)\int_0^1 du \,e^{i\,u\,p\cdot x}\, \phi_3^\sigma(u)\,, \\
	&\langle\,\pi(p)\,|\bar{q}(x)\,g_s\,\sigma_{\mu\nu}\,\gamma_5\,G_{\alpha\beta}(vx)\,q(0)\, |\,0\,\rangle \nn
	&=i\,f_{3\pi}\,\big(p_\alpha\,p_{[\mu}\,g_{\nu]\beta}^\perp-(\alpha\leftrightarrow\beta)\big)\,\int [d\alpha_i] e^{i\,\alpha_v p\cdot x} \,\Phi_3(\alpha_i)\,, \\
	&\langle\,\pi(p)\,|\bar{q}(x)\,g_s\,\gamma_\mu\,\gamma_5\,G_{\alpha\beta}(vx)\,q(0)\,|\,0\,\rangle \nn
	&=f_\pi \int [d\alpha_i]e^{i\,\alpha_v p\cdot x} \,\Big[\frac{p_{\mu} p_{[\alpha}x_{\beta]}}{p\cdot  x}\Phi_4(\alpha_i)  +p_{[\beta}g_{\alpha]\mu}^\perp  \Psi_4(\alpha_i)\Big]\,,\\
	&\langle\,\pi(p)\,|\bar{q}(x)\,g_s\,\gamma_\mu\,G^{\alpha\beta}(vx)\,q(0)\,|\,0\,\rangle \nn
	&=\frac{i}{2}\,f_\pi \epsilon^{\alpha\beta\sigma\tau}\,  \int [d\alpha_i]e^{i\,\alpha_qp\cdot x} \,\Big[\frac{p_{\mu} p_{[\sigma}x_{\tau]}}{p\cdot  x}\tilde{\Phi}_4(\alpha_i)  +p_{[\tau}g_{\sigma]\mu}^\perp  \tilde{\Psi}_4(\alpha_i)\Big]\,,
\end{align}
where the normalization parameter is defined as $\mu_\pi=m_\pi^2/(m_u+m_d)$, and  the integration measure over the momentum fractions reads
\begin{align}
	\int d[\alpha_i]=\int_0^1d\alpha_q\,d\alpha_{\bar{q}}\,d\alpha_g\, \delta(1-\alpha_q-\alpha_{\bar{q}}-\alpha_g)\,, \quad \alpha_v=\alpha_q+v\,\alpha_g\,.
\end{align}

It is worth noting that although the two-particle DAs $\phi_3^p(u)$, $\psi_4(u)$ and the three-particle DAs $\Phi_3(\alpha_i)$, $\Psi_4(\alpha_i)$, $\tilde{\Psi}_4(\alpha_i)$ are defined here for the theoretical completeness of the higher-twist pion DA basis, their  contributions either vanish exactly upon taking the Dirac trace or do not project onto the required antisymmetric tensor structure $\epsilon_{\mu\nu pq}$ of the correlation function. Consequently, they decouple completely and do not appear in our final sum rule evaluations in Eqs.~(\ref{2pht}) and (\ref{3pht}).

Based on the conformal expansion framework~\cite{Braun:1989iv}, the explicit expressions for the relevant higher-twist pion DAs utilized in our numerical evaluations are adopted from Ref.~\cite{Duplancic:2008ix}:
\begin{align}
	\phi_3^\sigma(u)&=6u\bar{u}\Big[1+\frac{5f_{3\pi}}{\mu_\pi f_\pi}\Big(1-\frac{\omega_{3\pi}}{10}\Big) C_2^{3/2}(u-\bar{u})\Big] \,,\\
	\phi_4(u)&=\frac{200}{3} \delta_\pi^2u^2\bar{u}^2 +8\delta_\pi^2\epsilon_\pi \Big[u\bar{u}(2+13u\bar{u}) +2u^3(10-15u+6u^2) \ln u \nn
	&+2\bar{u}^3(10-15\bar{u}+6\bar{u}^2) \ln\bar{u}\Big] \,,\\
	\Phi_4(\alpha_i)&=120\delta_\pi^2 \epsilon_\pi (\alpha_q -\alpha_{\bar{q}}) \alpha_q \alpha_{\bar{q}} \alpha_g \,,\\
	\tilde{\Phi}_4(\alpha_i)&=-120\delta_\pi^2  \alpha_q \alpha_{\bar{q}} \alpha_g \Big(\frac{1}{3} +\epsilon_\pi(1-3\alpha_g)\Big) \,.
\end{align}
Note that the implicit dependence of the higher-twist parameters and the distribution amplitudes on the renormalization scale has been omitted  for brevity.

\end{document}